# Three-dimensional X-ray visualization of axonal tracts in mouse brain hemisphere

Ryuta Mizutani[1,*], Rino Saiga[1], Masato Ohtsuka[2], Hiromi Miura[2], Masato Hoshino[3], Akihisa Takeuchi[3], and Kentaro Uesugi[3]

[1]Department of Applied Biochemistry, Tokai University, Hiratsuka, Kanagawa 259-1292, Japan

[2]Tokai University School of Medicine, Isehara, Kanagawa 259-1193, Japan

[3]Japan Synchrotron Radiation Research Institute (JASRI/SPring-8), Sayo, Hyogo 679-5198, Japan

*corresponding author: ryuta@tokai-u.jp

**ABSTRACT**

Neurons transmit active potentials through axons, which are essential for the brain to function. In this study, the axonal networks of the murine brain were visualized with X-ray tomographic microscopy, also known as X-ray microtomography or micro-CT. Murine brain samples were freeze-dried to reconstitute the intrinsic contrast of tissue constituents and subjected to X-ray visualization. A whole brain hemisphere visualized by absorption contrast illustrated three-dimensional structures including those of the striatum, corpus callosum, and anterior commissure. Axonal tracts observed in the striatum start from the basal surface of the cerebral cortex and end at various positions in the basal ganglia. The distribution of X-ray attenuation coefficients indicated that differences in water and phospholipid content between the myelin sheath and surrounding tissue constituents account for the observed contrast. A rod-shaped cutout of brain tissue was also analyzed with a phase retrieval method, wherein tissue microstructures could be resolved with up to 2.7 μm resolution. Structures of axonal networks of the striatum were reconstructed by tracing axonal tracts. Such an analysis should be able to delineate the functional relationships of the brain regions involved in the observed network.

**Introduction**

Neurons transmit active potentials through axons. Axons are sheathed with myelin, which is mainly composed of lipids such as phospholipids and glycolipids[1]. Axonal tracts of the corpus callosum and anterior and posterior commissures connect the left and right cerebral hemispheres and are essential for inter-hemisphere communication. It has been suggested that certain behavioral abnormalities are due to lateralization of brain function in the absence of inter-hemisphere communication[2]. Axonal tracts are also observed in the striatum constituted of the caudate and putamen nucleus. The striatum facilitates communication between the cerebral cortex and thalamus and is involved in motor control[3]. It has been reported that striatal activity is significantly altered in a number of pathologies, including Parkinson's disease and drug abuse[4]. It has also been suggested that the striatal circuit mediates a range of autism-associated behaviors, including social and cognitive deficits[5].

Although the three-dimensional network of axonal tracts in the human brain has been investigated with diffusion tensor imaging that measures the extent of water diffusion by observing proton magnetic resonance[6], the millimeter-order resolution of this method[7] imposes a limitation on visualizing microstructures of smaller axonal networks. Such cellular and subcellular structures can be visualized with X-ray tomographic microscopy, although soft tissues composed of light elements produce little contrast in X-ray images. In order to visualize soft tissue microstructures with X-rays, a number of methods have been reported for labeling tissue constituents with high atomic number (high-Z) elements[8]. High-Z probe labeling is primarily performed by immersing biological objects in high-Z reagent solutions so as to allow the probe to adsorb to the target structure. The staining efficacy depends on the ability of the reagent to penetrate the tissue. Since most staining procedures are performed under aqueous conditions, hydrophobic tissue components such as lipids act as barriers to probe permeation and thus hinder visualization.

Although high-Z probe labeling has been performed by introducing a probe from outside, X-ray visualization can also be performed by introducing a reporter gene that expresses a protein which interacts with X-rays. Since X-rays interact with high-Z elements efficiently, proteins that accumulate metals can be used as reporter proteins. We have reported X-ray tomographic visualization of bacterial cells overexpressing a metalloprotein, ferritin[9]. However, this method needs genetically engineered strains, which can only be prepared for well-established experimental organisms.

Since high-Z elements attenuate X-rays, high-Z probe labeling is not suitable for visualizing tissue samples with millimeter dimensions. Another way to visualize cellular structures is by detecting intrinsic X-ray contrasts of tissue components, such as contrasts between proteins and lipids. Since the myelin sheaths of axons are rich in phospholipids[1], axonal structures should be able to be visualized as lipid distributions. However, it is difficult to visualize such intrinsic differences even using phase contrast techniques, since water is the primary component of soft tissue by weight and hence the primary factor interacting with X-rays. The intrinsic X-ray contrast is masked by water infilling, thereby making it difficult to visualize tissue microstructures directly.

Freeze drying has been used in scanning electron microscopy for preparing soft tissue samples including those of brains[10]. It was reported that critical point drying modified the X-ray attenuation properties of tissue structures and facilitated visualization of a primate eye[11] and arthropod organs[12,13]. Therefore, the intrinsic X-ray contrast of brain tissue constituents should be visualized by removing water from the tissue.

In this study, murine brain samples were immersed in t-butyl alcohol and subjected to freeze drying so as to remove water from the tissue. A resin-embedded sample of cerebral tissue was also prepared as a control. The brain hemisphere samples were then visualized with X-ray tomographic microscopy, also known as microtomography or micro-CT, using monochromatic radiation from synchrotron sources. The obtained results indicated that three-dimensional microstructures of brain tissues can be visualized with X-ray by removing water from the tissue. An analysis of the attenuation coefficient distribution indicated that differences in water content are the primary factor affecting the observed X-ray contrast. The visualized structures of axonal tracts of striatum were traced in order to build a skeletonized wire model, indicating that striatum networks can be reconstructed with this method.

## Results

### Visualization of brain hemisphere

Freeze drying and critical point drying have been used for preparing samples for scanning electron microscopy[10]. The artefact possibly introduced by freeze drying is tissue shrinkage. In this study, we used t-butyl alcohol for freeze-drying brain samples, since it has been reported that shrinkage due to t-butyl alcohol dehydration prior to freeze drying is significantly less than that in ethanol in critical point drying[14]. Figure 1 and Supplementary Video S1 show the overall three-dimensional structure of a brain hemisphere sample prepared by freeze drying using t-butyl alcohol. This structure was visualized with absorption contrast microtomography using 12-keV monochromatic X-rays. The voxel width was 2.76 μm. Blood vessels were clearly visualized on the brain surface. Structures including those in the striatum were observed in a tomographic section (Figure 1c), indicating that the cerebral network can be visualized with this method. The results were replicated in hemisphere samples of three female and two male mice. Figure 2 shows paraffin sections of the right hemisphere. The height of the cerebral hemisphere in Figure 1c (4.1 mm) was approximately 10% smaller than that of a paraffin section of the contralateral hemisphere of the same mouse (Figure 2a; 4.6 mm). Although the tissue shrinkage caused ventricle dilatation, the overall architecture of the brain hemisphere had a structure comparable to that of the paraffin sections.

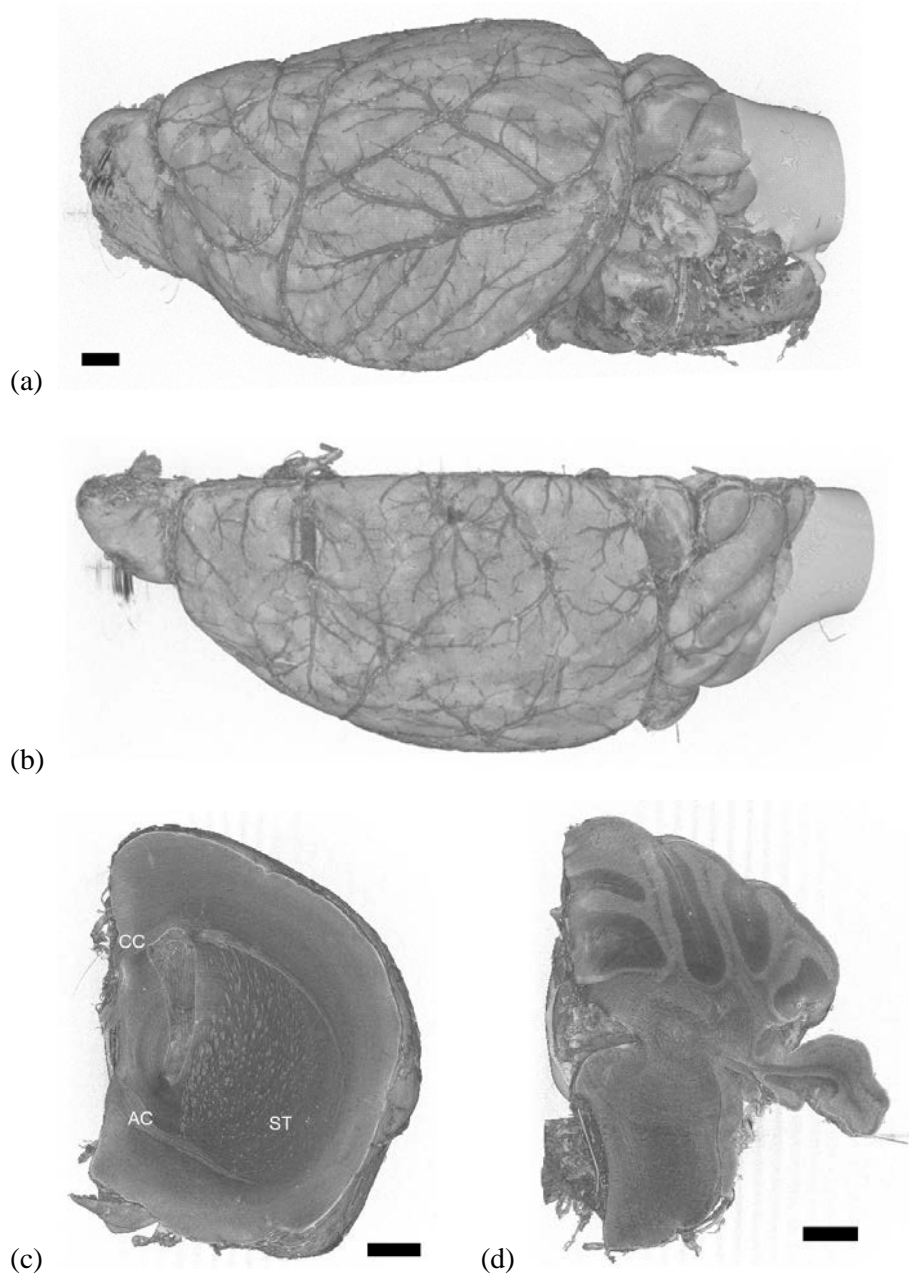

**Figure 1.** Three-dimensional rendering of left hemisphere of murine brain. Linear attenuation coefficients from 0.3 cm$^{-1}$ to 3.5 cm$^{-1}$ were rendered with gray scale. Scale bars: 500 μm. (a) Side and (b) top views of the entire hemisphere. (c) Coronal cutaway at mid-cerebrum. Corpus callosum (CC), anterior commissure (AC), and striatum (ST) structures appear as regions with high attenuation coefficients. (d) A coronal cutaway at the cerebellum. The high coefficient regions are granular layers. These are representative images of five hemisphere samples.

In a coronal section of the cerebrum (Figure 1c), the striatum network, the corpus callosum, and anterior commissure were observed as structures having higher X-ray attenuation coefficients compared with other brain regions. These anatomical structures correspond to tissue regions stained in a Klüver-Barrera section (Figure 2b), indicating that myelinated axons were visualized in the X-ray image. The number of axonal tracts observed in the microtomography section shown in Figure 1c is 473. The number of axonal tracts observed in the corresponding Klüver-Barrera section shown in Figure 2b is 492. The difference should owe to the position and direction of the sections. Therefore, these results indicate that almost all axonal tracts were visualized with X-ray. In a cerebellum section (Figure 1d), the granular layer was observed as high attenuation coefficient regions. Figure 3 shows a stereo rendering of the striatum network visualized in this sample. This figure was produced by rendering an attenuation coefficient range corresponding to the contrast of the striatum structure. Each tract starts from the basal surface of the cerebral cortex and ends at various positions in the basal ganglia.

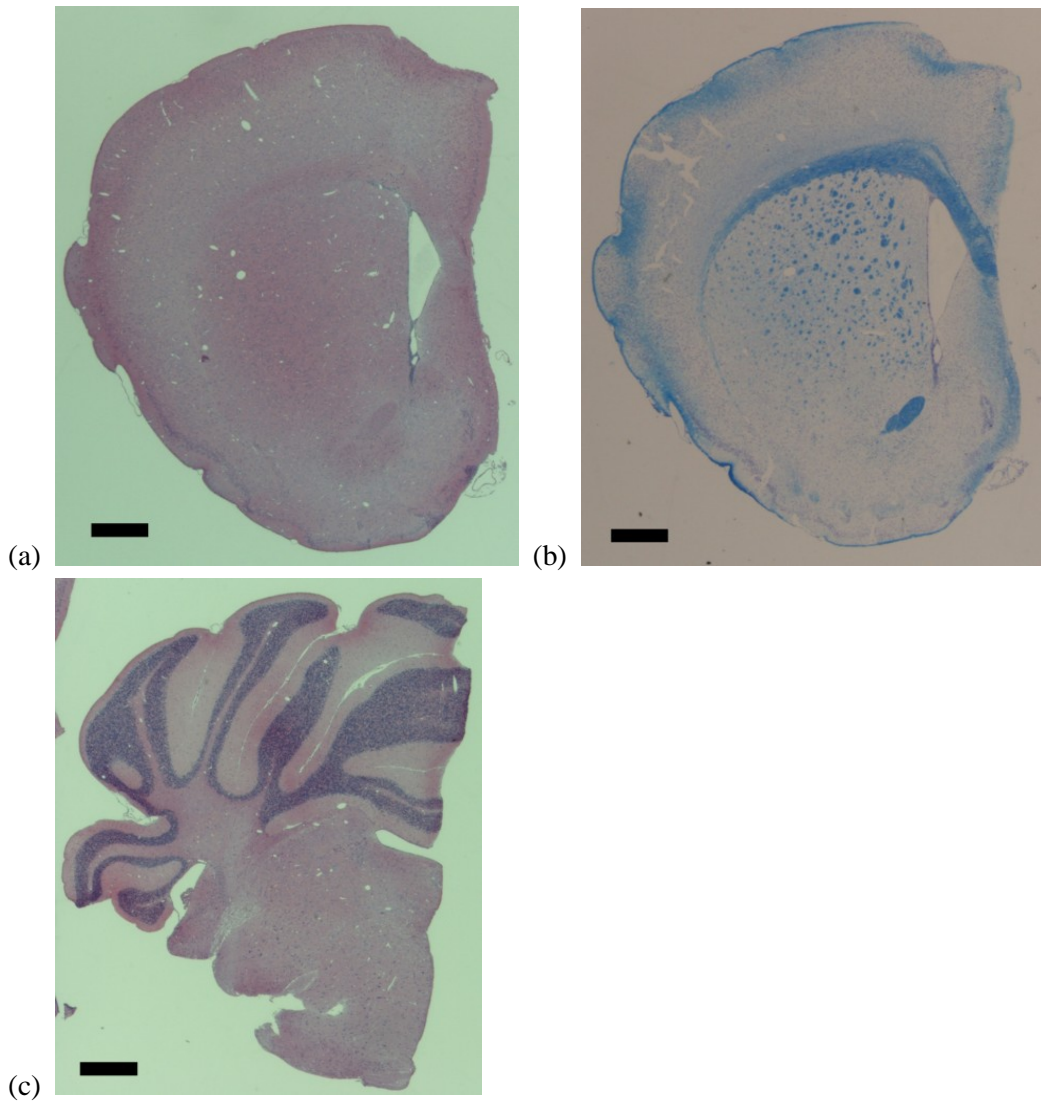

**Figure 2.** Paraffin sections of right hemisphere of murine brain. Sections 2 μm in thickness were prepared for hematoxylin-eosin staining, and 4 μm sections were prepared for Klüver-Barrera staining. Scale bars: 500 μm. (a) A hematoxylin-eosin section of cerebrum. The anterior commissure and corpus callosum appear as eosin-stained (pink) structures. (b) A Klüver-Barrera section of cerebrum. Myelinated axonal tracts in the anterior commissure, corpus callosum, and striatum appear as blue structures. (c) A hematoxylin-eosin section of cerebellum. Granular layers appear as hematoxylin-stained (purple) structures.

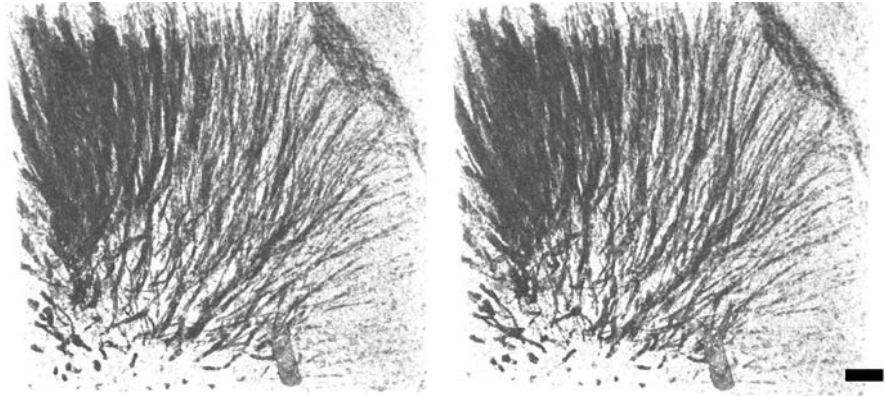

**Figure 3.** Stereo rendering of axonal tracts observed in the striatum of the hemisphere sample shown in Figure 1. The basal surface of the cerebral cortex can be seen at the upper right corner. A vessel can be seen at the bottom of the image. X-ray attenuation coefficients in a 1.1 × 1.1 × 0.46 mm$^3$ volume were rendered from 1.6 cm$^{-1}$ to 3.4 cm$^{-1}$. Dorsal is toward the top. Anterior view. Scale bar: 100 μm.

The three-dimensional distribution of X-ray attenuation coefficients (Figure 3) was traced in order to reconstruct the axonal network of striatum. Figure 4 shows stereo drawings of the skeletonized wire model of axonal network. The model was manually built by placing and connecting nodes in the three-dimensional image (Figure 4a) using a method reported in the analysis of human cerebral circuits[15]. It took only approximately 15 person-hours to build wire models in the three-dimensional image of $1.1 \times 1.1 \times 0.46$ mm$^3$ volume. Density picking and tracing functions implemented in dedicated software[16] enabled more rapid and efficient reconstruction of the axonal network compared with slice-by-slice registration of electron microscopy images. However, individual myelinated axons were not completely resolved at the micrometer resolution achieved in this study. Hence, the wire model reconstructed from the three-dimensional image represents tracts of axons, whereas axons bifurcating from one axonal tract and then converging into another tract were also observed. The obtained entire model shown in Figure 4b consists of 488 tracts. This nearly coincides with the number of axonal tracts observed in the microtomographic cross-section and the Klüver-Barrera section.

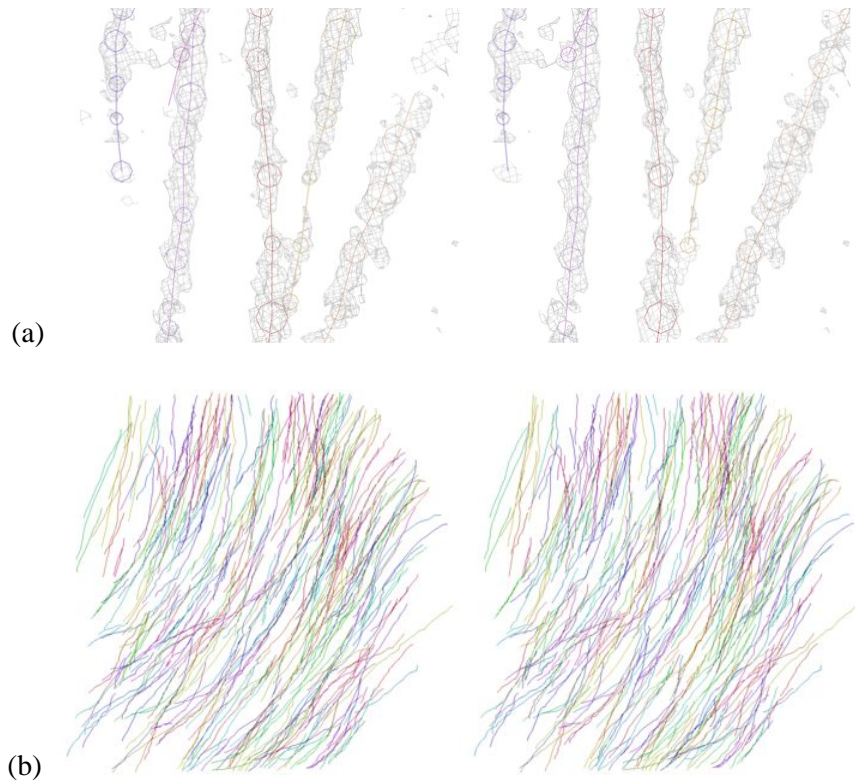

**Figure 4.** Stereo drawings of skeletonized wire model reconstructed by tracing axonal tracts. Skeletonized tracts are color-coded. (a) X-ray attenuation coefficient maps (gray) superposed on the wire model. The wire model was built by tracing these three-dimensional distributions of X-ray linear attenuation coefficients. Coefficient maps in a 2.76-μm grid are contoured at 3.7 cm$^{-1}$. Circles indicate the diameter assigned to each node. (b) Anterior view of the entire wire model of axonal tracts reconstructed in the region shown in Figure 3. The basal surface of the cerebral cortex is at the upper right corner. Dorsal is toward the top.

**Structures visualized with freeze-drying**

Rod-shape cerebral tissues were prepared by cutting tissue pieces along a direction perpendicular to the pial surface using razor blades. The obtained tissues were subjected to freeze drying or resin embedding and then visualized with 7.13-keV X-rays. Figure 5 shows tomographic sections of the rod-shaped samples. Fibrous networks were observed in the striatum of the freeze-dried sample (Figure 5a). The spatial resolution of this image was estimated to be 7.6 μm from a Fourier domain plot[17]. In contrast, no obvious structures were observed in a corresponding section of the resin-embedded sample (Figure 5b), indicating that the resin in the brain tissue hindered detection of the intrinsic contrast. Histograms of the X-ray linear attenuation coefficient of these tomographic sections are shown in Figure 5c. The attenuation coefficient of the freeze-dried sample had a broad profile ranging from the intensity of the axonal tracts to that of the background structures. In contrast, the profile of the resin-embedded sample had a narrow peak, indicating that the intrinsic tissue contrast observed in the freeze-dried sample was neutralized with the resin. The average attenuation coefficient of the resin-embedded sample was higher than that of the freeze-dried sample because of X-ray attenuation by the resin.

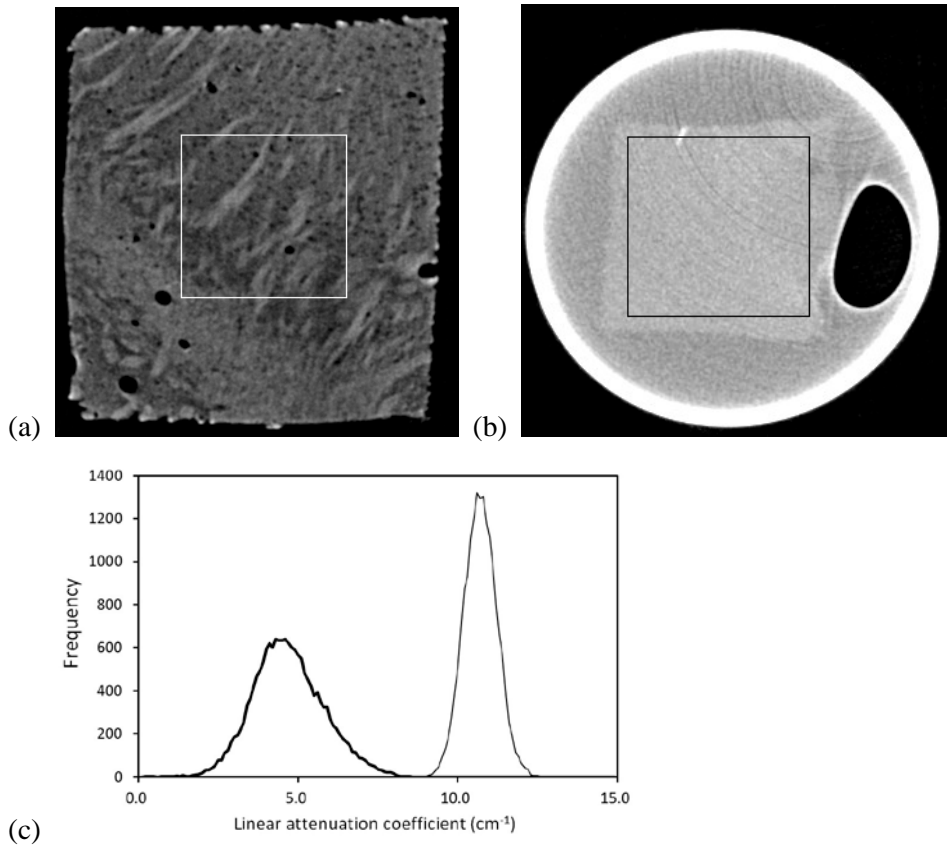

**Figure 5.** Tomographic sections of the striatum of freeze-dried and resin-embedded tissue samples. Linear attenuation coefficients from 0 cm$^{-1}$ to 15 cm$^{-1}$ are gray-scaled. Boxes with dimensions of 360 μm × 360 μm indicate the areas used for plotting the histograms. (a) Section of the freeze-dried sample. (b) Section of the resin-embedded sample. (c) Attenuation coefficient histograms. The number of pixels in each 0.1 cm$^{-1}$ bin is plotted against the attenuation coefficient. The thick line represents the results of the freeze-dried sample, and the thin line shows those of the resin-embedded sample.

The structure of the freeze-dried sample was also visualized using an image detector having 0.5-μm pixels. Finer structures can be visualized by taking images with finer voxels. However, the signal-to-noise ratio of the resultant three-dimensional image is proportional to three-halves the power of the voxel width. This is because photons per voxel are proportional to the third power of the voxel width, while the signal-to-noise ratio is proportional to the square root of the number of observation. Figure 6a shows a tomographic section visualized under conditions similar to those of Figure 5a but using a 0.5-μm pixel detector. Although its spatial resolution was estimated to be 1.6 μm from a Fourier domain plot[17], it is difficult to see axonal structures in this image. The same dataset was reconstructed by applying a phase retrieval method[18]. The obtained tomographic section is shown in Figure 6b. Axonal tracts and spherical low-density constituents of several micrometers in diameter are visible, although ring artefacts due to the tomographic reconstruction are also distinct. The diameter of these spherical structures coincides with those of cellular nuclei[19,20]. The spatial resolution of this image is estimated to be 2.7 μm, meaning that brain microstructures can be resolved with this method.

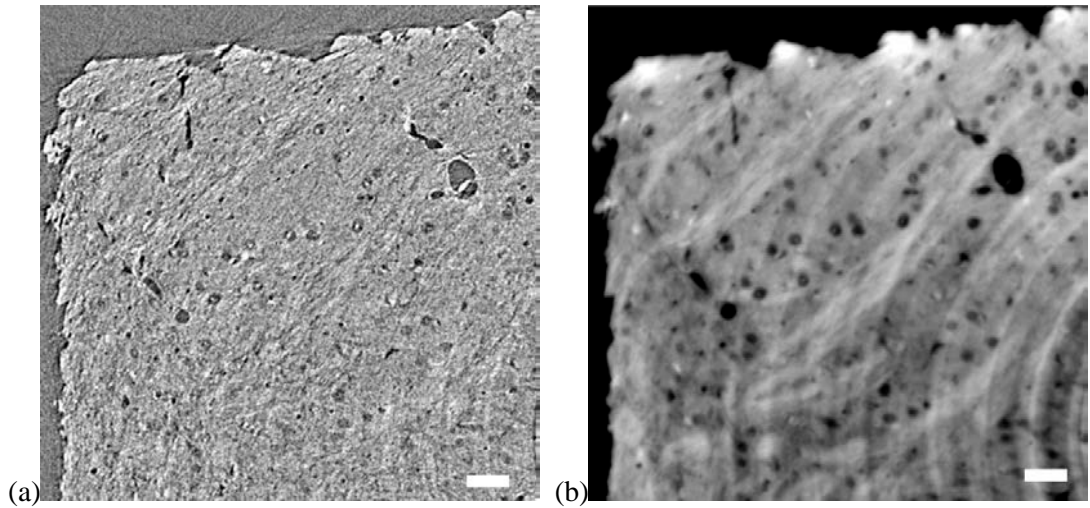

**Figure 6.** Tomographic sections of the striatum of the freeze-dried sample visualized with a 0.5-μm voxel width. These images approximately correspond to the upper left corner of the section shown in Figure 5a. Scale bars: 25 μm. (a) Absorption-contrast tomographic section. Linear attenuation coefficients from -15 cm$^{-1}$ to 15 cm$^{-1}$ are gray-scaled. (b) The same section reconstructed with the phase retrieval method. Arbitrary gray scale.

**Discussion**

The results of this study indicated that murine brain structures that cannot be observed in the resin-embedded sample were visualized in the freeze-dried sample. It has been suggested that the contrast change in a dried sample is due to the introduction of air interfaces in the tissue constituents[11]. Since the refractive index difference between the tissue constituents and the surrounding air is much larger than the difference between the tissue and water, the refraction enhancement had by the removal of the water can account for the contrast change. If this effect is the primary factor of visualization, the effect must be independent of the tissue constituents, including the cell bodies and myelinated axons. Although the refraction enhancement might still contribute to the visualization, only axonal structures and cellular nuclei were visualized in the freeze-dried sample in this study. The interface between the resin and tissue is not so obvious in Figure 5b. In contrast, axonal tracts are clearly discernible in Figure 5a, and edge overshoots between tissue constituents are not predominant. These results indicate that the refraction enhancement is not the primary factor of visualization. It has been reported that drying modified the X-ray attenuation properties of tissue of insect pupa[12] and that the X-ray contrast of arthropod tissues increased after drying the sample[13]. Therefore, it was suggested that some intrinsic difference in the X-ray interactions between one tissue constituent and the other should become evident as a result of removing water from the tissue.

Since the water removal visualized the tissue microstructure, the difference in water content may be the primary factor of the contrast. It has been reported that the water content of myelin is 40%[1], while that of overall brain tissue is approximately 80%[21]. The water content is almost independent of the animal species[21]. The contribution of water to the X-ray contrast observed in a freeze-dried sample can be estimated as follows. The X-ray mass attenuation coefficient of gray and white matter calculated by interpolating 6-keV and 8-keV coefficient data provided by ICRU[22] and NIST[23] is 16.7 cm$^2$/g at 7.13 keV, and that of water is 16.6 cm$^2$/g. The specific gravity of brain tissue[24] was reported to be 1.03 g/cm$^3$, giving an X-ray linear attenuation coefficient of brain tissue of 17.2 cm$^{-1}$. By taking account of the water content, the attenuation coefficient of the overall brain after removal of the water is calculated to be 3.9 cm$^{-1}$, while that of myelin is calculated to be 10.6 cm$^{-1}$. The histogram of attenuation coefficients of the freeze-dried sample (Figure 5) showed a broad peak ranging from 2.5 cm$^{-1}$ to 8.0 cm$^{-1}$. The maximum (4.3 cm$^{-1}$) almost coincided with the coefficient calculated for the overall brain. The higher part of the peak representing the attenuation coefficient of the axonal network is smaller than the coefficient calculated for myelin. This is because axonal tracts are composed of not only myelin but also other cellular and extracellular constituents, giving an average coefficient. Therefore, we suggest that the intrinsic water-content difference between tissue constituents mainly accounts for the X-ray contrast observed in the freeze-dried sample.

Since the myelin sheath of the axon is rich in phospholipids[1], the attenuation coefficient of the axonal structure can also be ascribed to phosphorus atoms. Phosphorus is a third period element in the periodic table and hence shows a larger X-ray interaction compared with light elements such as carbon and oxygen, i.e., those mainly composing the surrounding cells and extracellular matrix. It has been reported that dry myelin consists of 71% lipid by weight, of which 44% is phospholipid[1]. On the other hand, dry whole brain consists of 37% lipid by weight of which 58% is phospholipid[1]. By considering the specific gravity[24] and water content[1,21] of brain tissue, the phospholipid content of myelin is calculated to be 0.193 g/cm$^3$, while that of whole brain is 0.044 g/cm$^3$. The X-ray mass attenuation coefficient of phosphorus is 109.0 cm$^2$/g at 7.13 keV. If we assume the average molecular weight of phospholipids to be 750 Da, which approximately corresponds to C18 lipids reported for myelin[25], the linear attenuation coefficient due to phosphorus atoms is 0.87 cm$^{-1}$ for myelin and 0.20 cm$^{-1}$ for whole brain.

Therefore, the difference in the phosphorus distribution is not a determining factor but partly accounts for the X-ray contrast observed in the freeze-dried sample.

Such intrinsic differences were not visualized in the resin-embedded sample. Although resin embedding is appropriate for stabilizing micrometer- to nanometer-scale structures during acquisition of the X-ray images, the embedding procedure actually replaced the water molecules contained in the tissue with resin. The epoxy resin itself showed a linear attenuation coefficient of 9.0 cm$^{-1}$ at 7.13 keV. This is lower than that of water (16.6 cm$^{-1}$), though the X-ray attenuation by the resin should neutralize the contrast observed in the freeze-dried sample by resin infilling. Histograms calculated from upper-right and lower-left quarters of the tissue cross-section shown in Figure 5b gave similar peak profiles, suggesting that the peak width of the resin-embed sample is ascribed to the tissue structure and its image noise rather than ring artefacts. In contrast, the freeze-dried sample showed the broader profile (Fig. 5c) and has an advantage in its low overall X-ray attenuation compared with the resin-embedded sample. The low X-ray attenuation facilitates taking strong-intensity transmission images that can be used to make a precise determination of each pixel value. The low attenuation also allows visualization of the whole brain hemisphere to several millimeters in thickness.

A possible artefact introduced by freeze drying is the tissue deformation due to the shrinkage associated with the drying process. It has been reported that minor shrinkage artefacts in muscle tissue were detected in the X-ray image of a dried arthropod[13]. Ventricle dilatation possibly owing to tissue shrinkage was also observed in this study. Freeze drying is widely used for preparation of samples in scanning electron microscopy[10,14], and the shrinkage effect has been well-studied. Tissue deformation can be monitored by using alternative visualization methods, such as conventional light microscopy of paraffin sections. Since the axonal structure was only observed in the freeze-dried sample, we suggest that freeze drying should be applied to samples whose structure of interest is difficult to visualize with other methods.

X-ray microtomography allows three-dimensional visualization of tissue samples without destroying the sample. Conventional histology with light microscopy can visualize thin slice samples although its visualization depends on the slicing direction. This results in only a partial understanding of the three-dimensional structure. For example, a cross-section of tubular structure cannot be distinguished from that of sphere only from one two-dimensional image. In contrast, such structural features are easily discernible in one three-dimensional image. Since visible, infrared, or ultra-violet light do not pass through the brain hemisphere, three-dimensional structures of the whole hemisphere cannot be visualized even using confocal microscopy.

Three-dimensional tissue structures has been investigated with serial sectioning or serial block-face imaging[26,27]. However, thousands of sections or block faces must be prepared in order to visualize a millimeter-size sample, since electron beam cannot penetrate into the tissue. This sectioning procedure poses limitations on the total volume that can be analyzed. This is because correction of slice-by-slice deformations and registration of resultant images require a huge number of computations to reconstruct one three-dimensional structure from thousands of images. Therefore, it remains a painstaking task to analyze a number of samples to reveal differences between disease and normal control samples. In contrast, X-rays easily pass through biological samples, allowing three-dimensional visualization without sectioning. In this study, X-ray images of each sample were taken within 30 minutes. A three-dimensional image of the whole hemisphere consisting of over 6,000 tomographic slices was reconstructed within a few hours using a desktop PC. Since the sample is not destroyed upon the visualization, the analysis can be repeated by changing experimental conditions such as the X-ray energy in order to visualize specific elements or pixel width to visualize finer structures.

The primary limitation of the present method is the need for controlling biological hazards related to freeze-dried samples. Brain samples might contain infectious substances including prion proteins. Since freeze-dried samples can liberate flakes or dust of the tissue, biological hazards may accompany with the use of freeze-dried brain samples. If this method is applied to human tissues, biosafety concerns must be cleared prior to sample handling. Experiments even using animal samples should be designed by taking account of zoonotic infection hazards.

Another limitation of this method is that structures visualized with X-rays should be first identified by matching them with conventional histology. In this study, anatomical structures were identified with hematoxylin-eosin and Klüver-Barrera sections. However, once the structures are identified, microtomographic images can be used for structural studies, such as analysis of three-dimensional structural differences of brain networks using a number of samples.

Diffusion tensor imaging used for human brain tractography has been applied to experimental animals[28], though its millimeter-order resolution poses a difficulty as to the visualization of the axonal network of small animals. The results obtained in this study indicated that axonal tracts of the murine brain can be visualized at micrometer resolution by using X-ray microtomography. The network of axonal tracts can be reconstructed by building skeletonized wire models of the striatum as shown in Figure 4b. The network represents connections made by each axonal tract; hence, it is different from the conventional tractography in which tracts are built by tracing the direction of water diffusion. Therefore, the structure visualized in this study is a rather straightforward representation of a brain network compared with those obtained by tractography. Further analysis of this three-dimensional network should be conducted in order to reveal the innervation of each axonal tract and delineate the functional relationships of brain regions.

**Methods**

**Tissue sample preparation**

All mice were kept in the animal facility at Tokai University School of Medicine and fed ad libitum under a 12:12 light and dark cycle. C57BL/6J and BDF1 mice were purchased from CLEA Japan. Offspring mice were obtained by backcrossing a BDF2 mouse twice with C57BL/6J mice. All the animal experiments were approved by the Institutional Animal Care and Use Committee at Tokai University (permit number: 101041, 112004, 123012, 134013, 145001 and 151004) and performed in accordance with institutional guidelines.

The brains of the mice at eight weeks of age were dissected under perfusion fixation using formaldehyde saline. The right hemispheres of the dissected brains were subjected to paraffin embedding to prepare hematoxylin-eosin and Klüver-Barrera sections with the standard procedure. The left hemispheres were used for X-ray microtomographic analysis. Rod-shaped samples with approximate widths of 0.8 mm were excised from the left hemisphere using razor blades. The whole hemisphere and rod-shaped samples were dehydrated in an ethanol series and further soaked in t-butyl alcohol for 30 minutes for three times. The samples were frozen at -20ºC overnight and then lyophilized for 1 hour using a JFD-310 freeze drying device (JEOL, Japan). The resultant tissue samples were attached to mounting rods by using epoxy glue.

The resin-embedded samples were prepared as reported previously[15]. Resin embedding was performed by transferring the tissue sample first to ethanol, then to n-butyl glycidyl ether, and finally to Petropoxy 154 (Burnham Petrographics, ID, USA) epoxy resin. The brain tissues soaked in resin were cut into rod shapes with approximate widths of 0.5 mm under a stereomicroscope and transferred to a borosilicate glass capillary (W. Müller, Germany) filled with resin. The samples were kept at 90°C for 70 hours for curing the resin.

**X-ray microtomography**

X-ray microtomography with a large field of view was performed at the BL20B2 beamline[29] of the SPring-8 synchrotron radiation facility. The samples were mounted on the rotation stage using brass fittings. Transmission images were recorded with a CMOS-based X-ray imaging detector (AA40P and ORCA-Flash4.0, Hamamatsu Photonics, Japan) using a P43 scintillator screen. The field of view and effective pixel size of the detector are 5.65 mm × 5.65 mm and 2.76 µm × 2.76 µm, respectively. Transmission images of the hemisphere samples were acquired with a rotation step of 0.06° and an exposure time of 200 ms using 12-keV monochromatic X-rays. Images of the rod-shaped samples were acquired with a rotation step of 0.1° and an exposure time of 800 ms using 7.13-keV monochromatic X-rays. The X-ray energy was calibrated using a thin Fe plate.

X-ray microtomography at a higher resolution was performed at the BL20XU beamline[30] of SPring-8. The rod-shaped samples were mounted on the rotation stage by using brass fittings. Transmission images were recorded with a CMOS-based X-ray imaging detector (AA50 and ORCA-Flash4.0, Hamamatsu Photonics) using monochromatic radiation at 8 keV. X-ray energy was calibrated using Au foil. The LuAG scintillator screen of the detector was placed 7 mm from the sample. The viewing field and effective pixel size of the detector were 1.024 mm × 1.024 mm and 0.500 μm × 0.500 μm, respectively. Images were acquired with a rotation step of 0.1° and an exposure time of 150 ms. The data collection conditions are summarized in Table 1.

Absorption-contrast tomographic slices were reconstructed with a convolution-back-projection calculation using the RecView program[31] (available at https://mizutanilab.github.io/). Tomographic slices perpendicular to the sample rotation axis were reconstructed from the X-ray image datasets, giving the three-dimensional distribution of the linear attenuation coefficients. Phase retrieval[18] and subsequent tomographic reconstruction were performed using a program suite (available at http://www-bl20.spring8.or.jp/xct/index-e.html) specially built for data taken at SPring-8 beamlines. The spatial resolution of each slice was estimated from a Fourier domain plot[17]. Volume-rendered figures and a video of the obtained three-dimensional structures were prepared using the VG Studio MAX program (Volume Graphics, Germany) with the scatter HQ algorithm.

**Table 1.** Data collection conditions.

| Beamline | BL20B2 | BL20B2 | BL20XU |
|---|---|---|---|
| X-ray energy (keV) | 7.13 | 12.00 | 8.00 |
| Pixel size (μm) [a] | $2.76 \times 2.76$ | $2.76 \times 2.76$ | $0.500 \times 0.500$ |
| Pixel dynamic range (bits) | 16 | 16 | 16 |
| Viewing field size (pixels) [a] | $2,048 \times 2,048$ | $2,048 \times 2,048$ | $2,048 \times 2,048$ |
| No. of sample frames per dataset | 1,800 | 3,000 | 1,800 |
| Degrees per frame | 0.10 | 0.06 | 0.10 |
| Exposures per frame (ms) | 800 | 200 | 150 |
| Dataset collection time (s) | 1,580 | 780 | 350 |

[a] Width $\times$ height

**Model building**

A three-dimensional region of $1.1 \times 1.1 \times 0.46$ mm$^3$ of the obtained image corresponding to the mid-striatum was subjected to model building[15] by using the program MCTrace[16] (available from https://mizutanilab.github.io/). Neuronal processes were traced by tracking the attenuation coefficient distribution down to 3.3 cm$^{-1}$. This is equivalent to 6.0 times the standard deviation (6.0 σ) of attenuation coefficients in the volume subjected to the model building. The model building was performed using a method like those reported for crystallographic studies of macromolecular structures[15]. Node coordinates of axonal tracts were placed and connected while the three-dimensional map of the attenuation coefficients was viewed.

A set of nodes linked through connections was designated as a "trace" and numbered serially. Each node belonging to a certain trace was discriminated using node names. The coordinate system of the model was defined so as to place the -x/+x axis nearly along the dorsoventral direction, the -y/+y axis along the left-to-right direction, and the -z/+z axis along the anteroposterior direction. The coordinate origin was placed at the origin of the coefficient map. Cartesian coordinates of model nodes given in micrometers were enumerated in Protein Data Bank format and posted in the microtomographic structure repository on the institution's web page (https://mizutanilab.github.io/).

**Acknowledgements**

We thank Akemi Kamijo and Sanae Ogiwara (Support Center for Medical Research and Education, Tokai University) for assistance in raising the mice. We thank Hideo Tsukamoto, Masayoshi Tokunaga and Noboru Kawabe (Support Center for Medical Research and Education, Tokai University) for assistance in preparing the tissue samples. We thank Kiyoshi Hiraga (Technical Service Coordination Office, Tokai University) for assistance in preparing the brass fittings for mounting the sample. This work was supported in part by Grants-in-Aid for Scientific Research from the Japan Society for the Promotion of Science (nos. 21611009, 25282250, and 25610126). The synchrotron radiation experiments were performed at SPring-8 with the approval of the Japan Synchrotron Radiation Research Institute (JASRI) (proposal nos. 2013B0034, 2013A1441, 2014B1096, 2015A1160, and 2015B1101).

**Author contributions**

R.M., R.S., M.O., and H.M. prepared the brain tissue samples. R.M., R.S., M.H., A.T., and K.U. performed the synchrotron radiation experiments. R.M. and R.S. analyzed the data. R.M. designed the study and wrote the manuscript. All authors contributed to the data interpretation and reviewed the manuscript.

**Competing financial interests**

The authors declare no competing financial interests.

**Supplementary Video S1.**

(https://youtu.be/J87oHafsONA)

Three-dimensional rendering of the left hemisphere of murine brain. Linear attenuation coefficients from 0.3 cm$^{-1}$ to 3.5 cm$^{-1}$ were rendered with gray scale. First, the entire hemisphere is rotated along the sagittal axis to show its overall structure. Then, coronal cutaways are shown by moving the section plane from the olfactory bulb to the medulla. Subsequently, the section plane goes back and the entire hemisphere returns to the original position.